# Evaluation of the Feasibility of Phosphorene for Electronic DNA Sequencing Using Density Functional Theory Calculations

**Matthew B. Henry**

Department of Engineering and Physics, University of Central Oklahoma, Edmond,OK 73034

**Mukesh Tumbapo**

Department of Engineering and Physics, University of Central Oklahoma, Edmond,OK 73034

**Kolby Wilson**

Department of Engineering and Physics, University of Central Oklahoma, Edmond,OK 73034

**Benjamin O. Tayo**

Department of Engineering and Physics, University of Central Oklahoma, Edmond,OK 73034

**Abstract:** Electronic DNA sequencing using two-dimensional (2D) materials such as graphene has recently emerged as the next-generation of DNA sequencing technology. Owing to its commercial availability and remarkable physical and conductive properties, graphene has been widely investigated for DNA sequencing by several theoretical and experimental groups. Despite this progress, sequencing using graphene remains a major challenge. This is due to the hydrophobic nature of graphene, which causes DNA bases to stick to its surface via strong $\pi - \pi$ interactions, reducing translocation speed and increasing error rates. To circumvent this challenge, the scientific community has turned its attention to other 2D materials beyond graphene. One such material is phosphorene. In this article, we performed first-principle computational studies using density functional theory (DFT) to evaluate the ability of phosphorene to distinguish individual DNA bases using two detection principles, namely, nanopore and nanoribbon modalities. We observe that binding energies of DNA bases are lower in phosphorene compared to graphene. The energy gap modulations due to interaction with DNA bases are very significant in phosphorene compared to graphene. Our studies show that phosphorene is superior to graphene, and hence a promising alternative for electronic DNA sequencing.

## Introduction

The progress towards cheaper and faster sequencing has been very impressive since the Human Genome Project first sequenced the human genome using the classical Sanger method (Lander et al. 2001). Recently, Oxford Nanopore Technologies developed a sequencing device based on protein nanopores (Mikheyev and Tin 2014). Despite this progress, there are still several challenges (Jain et al. 2015, Dekker 2007, Heerema and Dekker 2016) with DNA sequencing using protein nanopores such as: high startup and consumables costs; sensitivity of pore to environmental conditions *e.g.,* temperature, pH, and applied voltage; and high error rate (~15%). Due to these challenges, the current focus is on electronic DNA sequencing using 2D materials. Electronic DNA sequencing has the capability to produce larger current signals (~ μA) compared to the low ionic-current signals (~ pA) used in protein nanopores (Heerema 2016). Electronic methods could also lead to label-free, single-nucleotide, long read length automated sequencing without requiring additional consumables (Patel et al. 2017), which could drive down cost and improve accuracy.

Owing to its commercial availability and remarkable physical and conductive properties, graphene has been widely investigated for DNA sequencing by several theoretical (Prasongkit et al. 2011 and Prasongkit et al. 2015) and experimental (Schneider et al. 2010, Merchant et al. 2010, and Garaj et al. 2010) groups. Despite this progress, sequencing using graphene remains a major challenge. This is due to the hydrophobic nature of graphene, which causes DNA bases to stick to its surface via $\pi - \pi$ interactions, reducing translocation speed and increasing error rates (Sathe et al. 2011 and Wells et al. 2012).

Recently, the scientific community has turned its attention to other 2D materials beyond graphene. For instance, molybdenum disulfide ($MoS_2$) has been shown to produce better signal-to-noise ratios, and non-stickiness of DNA bases to its surface (Farimani et al. 2014). Furthermore, the presence of a finite energy gap (energy difference between valence band maximum and conduction band minimum) makes $MoS_2$ to be suitable for advanced sequencing devices such as field-effect transistors (Graf et al. 2019).

Another potential 2D material for DNA sequencing is phosphorene (Novoselov et al. 2016). Phosphorene is an elemental 2D material similar to graphene with remarkable electronic properties including a finite energy gap. Furthermore, phosphorene is hydrophilic and biocompatible (not toxic to cells), making it suitable for biological applications (Cortes-Arriangada et al. 2018, Kumawat et al. 2019, and Kumawat and Pathak 2019).

In this article, we performed first-principle computational studies using density functional theory (DFT) to evaluate the ability of graphene and phosphorene to distinguish individual DNA bases using two detection principles, namely, nanopore and nanoribbon modalities.

## Materials and Methods

We focus on two advanced detection principles, namely, the nanopore and nanoribbon methods (Heerema et al. 2016), as shown in Figure 1. For simplicity, we will refer to our nanodevice concepts using the following abbreviations: GNP (graphene nanopore), PNP (phosphorene nanopore), GNR (graphene nanoribbon), and PNR (phosphorene nanoribbon). The four DNA bases are guanine (G), adenine (A), cytosine (C), and thymine (T). For GNP, the active region has a dimension of 1.91 nm x 1.80 nm, with a pore diameter of 1.07 nm. For PNP, the active region measures 2.48 nm x 1.20 nm, with a pore diameter of 1.03 nm. For GNR, the dimension of the active region is 1.07 nm x 1.17 nm. For PNR, the dimensions are 1.33 nm x 1.32 nm. For both nanopore and nanoribbon systems, the size of the active region is comparable to the interbase distance ~0.7 nm (Lagerqvist et al. 2006), and hence suitable for single-base resolution. For the nanoribbon model, the DNA bases were placed 3.0 Å above center of the nanoribbon prior to geometry optimization calculations. The structural relaxation calculations were performed at the B3LYP level of theory using the 6-31G (d, p) basis set, with a force convergence cutoff of 0.02 eV/Å (Kumawat et al. 2019). All calculations were performed using the GAUSSIAN 16 software package (Frisch et al. 2016). Computational resources were provided by the University of Central Oklahoma Buddy Supercomputing Center.

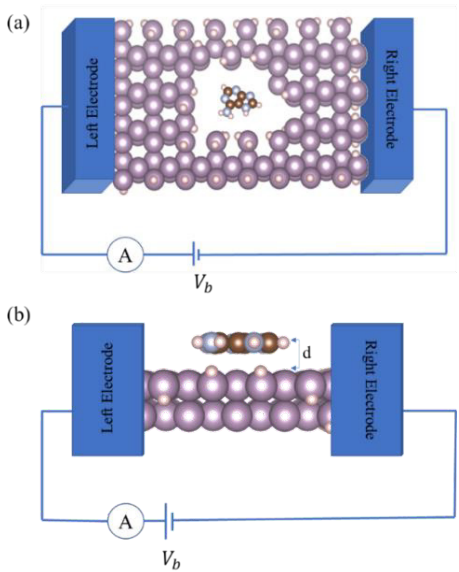

**Figure 1.** Schematic of electronic DNA sensing device concepts. (a) Translocation of DNA base through nanopore causes variations in in-plane current. (b) Changes in electronic current due to physisorption of DNA bases onto surface of 2D nanoribbon can be detected.

To evaluate the ability of graphene and phosphorene to distinguish individual DNA bases, two evaluation metrics were computed. The binding energy was calculated for both nanopore and nanoribbon methods as the difference in total energy, that is, $E_b = E_{system+base} - E_{system} - E_{base}$. The energy gap was calculated as the difference between the HOMO and LUMO energies: $E_{gap} = E_{HOMO} - E_{LUMO}$.

## Results and Discussions

Table 1 shows the energy gaps and binding energies for the four systems considered. The binding energies for GNP (0.871 – 1.063 eV) are larger than those for PNP (0.207 – 0.405 eV). Similarly, the binding energies for GNR (0.423 – 0.592 eV) are larger than those for PNR (0.182 – 0.330 eV). For PNR, our calculated binding energy order (G > A > C > T) is consistent with reported results using nanoribbons from $MoS_2$ (Farimani et al. 2014) and phosphorene (Kumawat and Pathak, 2019). Figure 2 shows the binding energy comparisons for all four systems. It shows that graphene nanomaterials typically have a higher binding energy, and hence greater tendency for bases to stick their surfaces, compared to phosphorene nanomaterials.

To further compare the performance of each system, we computed the change in energy gap as follows: $\Delta E_{gap} = E_{gap}(system + base) - E_{gap}(system)$. Table 2 shows the magnitude of energy gap change for all systems. For GNP, the change in energy gap is small ($\Delta E_{gap} \sim 7 - 10$ meV), while for PNP, the change is very significant ($\Delta E_{gap} \sim 13 - 281$ meV). Similarly, for GNR, the change in energy gap is small ($\Delta E_{gap} \sim 1 - 3$ meV), while for PNR, the change is very significant ($\Delta E_{gap} \sim 16 - 358$ meV).

Our idealized calculations show that phosphorene is superior to graphene for electronic DNA sequencing. In our model calculations, the effect of ions and solvating water were not included. We would expect the effects of solution and orientation of bases to produce changes in the magnitude of the calculated quantities, but not the trends (Henry et al. 2021, Feliciano et al. 2018, and Lagerqvist et al. 2007).

**Table 1.** Energy Gap and Binding Energy (absolute value) for GNP, PNP, GNR, and PNR.

| Base | Energy Gap (eV) | | | | Binding Energy (eV) | | | |
| --- | --- | --- | --- | --- | --- | --- | --- | --- |
| | GNP | PNP | GNR | PNR | GNP | PNP | GNR | PNR |
| Pristine | 0.221 | 3.070 | 0.259 | 3.038 | - | - | - | - |
| G | 0.228 | 3.083 | 0.260 | 2.680 | 0.888 | 0.395 | 0.592 | 0.330 |
| A | 0.230 | 2.789 | 0.257 | 2.783 | 0.936 | 0.307 | 0.546 | 0.293 |
| C | 0.230 | 3.046 | 0.262 | 3.022 | 1.063 | 0.405 | 0.578 | 0.182 |
| T | 0.231 | 3.025 | 0.258 | 3.055 | 0.871 | 0.207 | 0.423 | 0.169 |

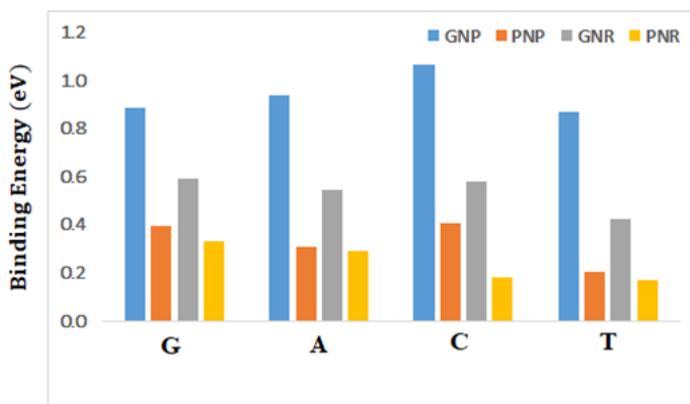

**Figure 2.** Binding energy for graphene and phosphorene device concepts.

**Table 2.** Change in energy gap for graphene and phosphorene models.

| Base | $\Delta E_{gap}$ (eV) | | | |
| --- | --- | --- | --- | --- |
| | GNP | PNP | GNR | PNR |
| G | +0.007 | +0.013 | +0.001 | -0.358 |
| A | +0.009 | -0.281 | -0.002 | -0.255 |
| C | +0.009 | -0.024 | +0.003 | -0.016 |
| T | +0.010 | -0.045 | -0.001 | +0.017 |

In conclusion, using two metrics, namely the binding energy and energy gap, we calculated the modulation of electronic properties of nanomaterials from graphene and phosphorene due to interaction with DNA bases using two advanced detection principles, namely, nanoribbon and nanopore concepts. Our calculations show that the binding energies for phosphorene systems are generally lower compared to graphene. Also, the modulation in energy gaps are pretty significant for phosphorene nanomaterials compared to graphene. Our studies show that phosphorene is superior to graphene, and hence a promising alternative to graphene for electronic DNA sequencing applications.

## Acknowledgments

Funding for this research was provided by the Office of Research and Sponsored Programs at the University of Central Oklahoma and the CURE-STEM grant from the College of Mathematics and Sciences.